\documentclass[aps,prd,twocolumn,superscriptaddress,showpacs,preprintnumbers]{revtex4-1}
\usepackage[colorlinks=true, pdfstartview=FitV, linkcolor=red, citecolor=blue, urlcolor=black, pdftitle={},pdfauthor={Tomoya Hayata},pdfsubject={}, pdfkeywords={}]{hyperref}
\usepackage[pdftex]{graphicx}
\usepackage{setspace,bm}
\usepackage{latexsym,amssymb,amsmath,mathrsfs}
\usepackage{color}
\newcommand{\cH}{{\cal H}}

\newcommand{\cO}{{\cal O}}

\newcommand{\vphi}{\varphi}

\newcommand{\ve}{\varepsilon}

\newcommand{\be}{\begin{equation}}      
\newcommand{\ee}{\end{equation}}      
\newcommand{\bea}{\begin{eqnarray}}      
\newcommand{\eea}{\end{eqnarray}}

\newcommand{\im}{\mathrm{i}}

\begin{document}
\title{Anomalous transport phenomena from dissipative charge pumping}

\author{Tomoya Hayata}
\affiliation{
Department of Physics, Chuo University, 1-13-27 Kasuga, Bunkyo, Tokyo, 112-8551, Japan 
}
\email[]{hayata@phys.chuo-u.ac.jp}

\date{\today}

\begin{abstract}

The Berry curvature involving time and momentum derivatives, which we term emergent electric field, induces a nondisspative current known as the adiabatic charge pumping or Thouless pumping in periodically driven systems. 
We study dissipative currents originated from the interplay between emergent electric fields and electric/magnetic fields in two and three dimensions on the basis of the Boltzmann transport theory. 
As an example of two-dimensional models, we study the Rashba Hamiltonian with time-dependent and anisotropic spin-orbit coupling. 
We show that the interplay between emergent electric fields and electric fields leads to a current transverse to electric fields, which is symmetric and contributes to the entropy production.
As an example of three-dimensional models, we study the Weyl Hamiltonian under AC electric fields. 
We show that the interplay between emergent electric fields and magnetic fields leads to a Hall-type current at zero DC electric fields, which is now transverse to DC magnetic fields: 
$j_{x}=\sigma_{xy}B_y$ ($\sigma_{xy}=-\sigma_{yx}$).
The Hall photocurrent is relevant in the inversion symmetry breaking Weyl semimetals such as TaAs or SrSi$_2$.
\end{abstract}

\maketitle

\section{Introduction} 
Transport phenomena under external fields such as electromagnetic fields, temperature  gradient, and mechanical strain are one of the fundamental subjects in physics.
Although familiar and classical effects such as Ohm's law, Hall effect, Seebeck effect, and Nernst effect have been found in the 19th century,
new transport effects such as quantum Hall effect~\cite{PhysRevLett.45.494,PhysRevB.23.5632,PhysRevLett.49.405}, 
(quantum) anomalous Hall effect~\cite{PhysRev.95.1154,PhysRevLett.88.207208,PhysRevLett.93.206602,Sinitsyn,RevModPhys.82.1539,RevModPhys.82.1959,Yu61,Chang167}, 
(quantum) spin Hall effect~\cite{PhysRevLett.83.1834,PhysRevLett.92.126603,Kato1910,PhysRevLett.94.047204,2006Natur.442..176V,PhysRevLett.95.226801,Bernevig1757} 
and chiral magnetic effect (CME)~\cite{NIELSEN1983389,PhysRevD.78.074033,PhysRevB.88.104412,PhysRevX.5.031023} 
still have been found in last four decades in systems with breaking of time-reversal symmetry and/or inversion symmetry (parity).

It was soon realized that those new transport phenomena can be uniformly described by using the Berry phase and Berry curvature~\cite{Berry45}.
We can define the Berry connection $\bm a=u^\dagger i\nabla_p u$ and Berry curvature $\bm b=\nabla_p\times \bm a$ in (three-dimensional) momentum space, 
by using wave function $u(\bm p)$ [$\bm p$ are monetum].
We term the Berry curvature $\bm b$ emergent magnetic field, 
which leads to a momentum analogue of Lorentz force (anomalous group velocity) in the semiclasscial wave packet dynamics of electrons~\cite{PhysRevB.59.14915,RevModPhys.82.1959}.
The aforementioned transport phenomena are originated from the anomalous group velocity because of the interplay between the emergent magnetic field $\bm b$, 
and electric fields $\bm E$ or magnetic fields $\bm B$, e.g., $\bm j_{\rm Hall}\sim \bm b\times \bm E$, and $\bm j_{\rm cme}\sim \bm B(\bm v\cdot \bm b)$, 
where $\bm v=\nabla_p\ve$ is the group velocity of Bloch electrons. 

When Hamiltonian depends on time $t$ under time-dependent external fields such as AC electric fields, wave function depends on $t$ [$u(t,\bm p)$]. 
Now we can define time-component of the Berry connection $a_0=u^\dagger i\partial_t u$, and the associated Berry curvature 
$f_{p_it}=\partial_{p_i}a_0-\partial_t a_{p_i}$ ($p_{i=x,y,z}$), which behaves as electric field in momentum space [$\bm e=(f_{p_x t},f_{p_y t},f_{p_y t})$],
and induces an analogue of the coulomb force in the semiclasscial wave packet dynamics of electrons~\cite{PhysRevB.59.14915,RevModPhys.82.1959}.
We term the Berry curvature $\bm e$ emergent electric field.
It is known that the emergent electric field induces a nondissipative DC current in periodically driven systems, 
which is referred to as the adiabatic charge pumping or Thouless pumping~\cite{PhysRevB.27.6083,PhysRevB.47.1651,RevModPhys.66.899,Nakajima,Lohse}.
However in contrast to the emergent magnetic field $\bm b$, the transport phenomena originated from  
the interplay between it and other external fields such as electromagnetic fields have not been discussed yet.

In this paper we study dissipative currents originated from the interplay between emergent electric fields and electric/magnetic fields on the basis of the Boltzmann transport theory with the relaxation time approximation.
Although we can consider a natural analogue of the anomalous Hall effect, that is,  the Hall current induced by emergent electric fields along the direction transverse to them in the presence of magnetic fields, $\bm j_{\rm Hall}\sim \bm B\times \bm e$~\cite{PhysRevB.95.125137}, it flows in momentum space and is not relevant in transport phenomena in real space.
To discuss the effect of the interplay in real-space transport phenomena, we need to consider the change of the distribution function by external fields and study dissipative currents.
We show, by studying the Rashba and Weyl Hamiltonians as examples, that there arise new types of anomalous currents originated from the interplay between emergent electric fields and electric/magnetic fields.
Below we use natural units ($\hbar=c=1$).

\section{Two dimensional system}
\subsection{Model Hamiltonian}
\label{sec:rashba}

We consider the two-dimensional fermion model with spin-orbit and Zeeman couplings. 
The model Hamiltonian is given as
\be
\cH =\frac{p_x^2+p_y^2}{2m}+\sigma_x \lambda_y(t)p_y-\sigma_y \lambda_x(t)p_x-M(t)\sigma_z,
\label{eq:Rashba1}
\ee
where $\sigma_i$ are the $2\times2$ Pauli matrices. 
$\lambda_x(t)$ and $\lambda_y(t)$ are anisotropic coupling strengths of the Rashba spin-orbit interaction~\cite{Rashba,0022-3719-17-33-015} in surface states with $C_{2v}$ point group symmetry~\cite{0953-8984-21-9-092001,PhysRevB.81.235438}. Both of them depend on the potential gradient perpendicular to the surface, 
$\lambda_{x,y}\sim \partial_z V$~\cite{PhysRevB.81.235438} (We term the direction perpendicular to the system $z$ direction), 
and are experimentally controlled by applying gate voltage~\cite{PhysRevLett.78.1335}.
We assume that $\lambda_x(t)$ and $\lambda_y(t)$ periodically oscillate with the same frequency $\omega$, 
which is physically realized by oscillating gate voltage with the frequency $\omega$ e.g., $V\sim V_0(z)\cos\omega t$.
$M(t)$ is the coupling between electron's spin and external magnetic field or (mean-field) magnetization.
We also assume that $M(t)$ depends on time, which is physically realized by time-dependent external magnetic field perpendicular to the system.
We rewrite Eq.~\eqref{eq:Rashba1} as
\be
\cH =R_0\bm 1+R_i\sigma_i,
\label{eq:Rashba2}
\ee
where $R_0=(p_x^2+p_y^2)/2m$, $\bm R(\bm p)=(R_x,R_y,R_z)=(\lambda_y p_y,-\lambda_x p_x,-M)$, and $\bm 1$ is the $2\times2$ unit matrix.
We hereafter employ the Einstein convention for repeated indices. 
The instantaneous eigenvalues of the Rashba Hamiltonian~\eqref{eq:Rashba1} or~\eqref{eq:Rashba2} are given as 
\bea
\varepsilon_{\pm} &=&R_0\pm R(\bm p) 
\nonumber \\
&=&\frac{p_x^2+p_y^2}{2m}\pm\sqrt{\lambda_x^2p_x^2+\lambda_y^2p_y^2+M^2} ,
\label{eq:Rashba_eigen}
\eea
where $R(\bm p)=|\bm R(\bm p)|$.
The instantaneous eigenvectors read
\be
u^{-}(t,\bm p)=  
\begin{pmatrix}
\sin\frac{\theta}{2}e^{-i\vphi} \\
-\cos\frac{\theta}{2}
\end{pmatrix},\;
u^{+}(t,\bm p)=  
\begin{pmatrix}
\cos\frac{\theta}{2}e^{-i\vphi} \\
\sin\frac{\theta}{2}
\end{pmatrix},
\label{eq:Rashba_wave}
\ee
where $R$, $\theta$ and $\varphi$ are the spherical coordinates in $\bm R$ space.
Then we can obtain the Berry connection and Berry curvature in $\bm R$ space as
\bea
\bm a^{\pm}_R &=&\left(u^{\pm}\right)^\dagger\im\nabla_R u^{\pm}= 
\frac{1\mp\cos\theta}{2R\sin\theta}\hat{\vphi} , \\
\bm b_R &=&\nabla_R\times\bm a^{\pm}_R= 
\pm\frac{\hat{R}}{2R^2} ,
\eea
where $\hat{R}$, $\hat{\theta}$, and $\hat{\vphi}$ are the unit vectors in the spherical coordinates.
For later purpose, we calculate the Berry curvature in three-dimensional space $\xi_\mu=(t,p_x,p_y)$:
\bea
a^{\pm}_{\xi_\mu} &=&\left(u^{\pm}(p)\right)^\dagger\im\partial_{\xi_\mu} u^{\pm}(p) , \\
f^{\pm}_{\xi_\mu\xi_\nu} &=& \partial_{\xi_\mu}a^{\pm}_{\xi_\nu}-\partial_{\xi_\nu}a^{\pm}_{\xi_\mu}.
\eea
We can obtain $f^{\pm}_{\xi_\mu\xi_\nu}$ by using the pullback from $\bm R$ space to $\bm\xi$ space:
\begin{align}
\begin{split}
f_{p_xp_y}^{\pm} 
&= \partial_{p_x} R^m\partial_{p_y} R^l\epsilon^{mln}b_R^{\pm n}
\\
&= \mp\frac{1}{2}\frac{\lambda_x\lambda_yM}{\left(\lambda_x^2p_x^2+\lambda_y^2p_y^2+M^2\right)^{\frac{3}{2}}} ,
\end{split}
\label{eq:Berry_curvature1}
\end{align}
\begin{align}
\begin{split}
\int_0^{\frac{2\pi}{\omega}} \frac{\omega dt}{2\pi} f_{p_xt}^{\pm} 
&= \int_0^{\frac{2\pi}{\omega}} \frac{\omega dt}{2\pi}\partial_{p_x} R^m\partial_{t} R^l\epsilon^{mln}b_R^{\pm n}
\\
&=\mp \int_0^{\frac{2\pi}{\omega}} \frac{\omega dt}{2\pi}\frac{1}{2}\frac{p_y\lambda_x\left(\partial_{t} \lambda_y M- \lambda_y \partial_{t} M\right)}{\left(\lambda_x^2p_x^2+\lambda_y^2p_y^2+M^2\right)^{\frac{3}{2}}} 
\\
&\sim\mp \frac{1}{4}\frac{\omega p_y\left(\lambda_{x0}\Delta_y-\lambda_{y0}\Delta_x\right) \Delta_M\cos\delta}{\left(\lambda_{x0}^2p_x^2+\lambda_{y0}^2p_y^2+M_0^2\right)^{\frac{3}{2}}}
\\
&\pm \frac{3}{4}\frac{\omega p_y\lambda_{x0}^2p_x^2\left(\lambda_{x0}\Delta_y-\lambda_{y0}\Delta_x\right) \Delta_M\cos\delta}{\left(\lambda_{x0}^2p_x^2+\lambda_{y0}^2p_y^2+M_0^2\right)^{\frac{5}{2}}},
\label{eq:Berry_curvature2}
\end{split}
\\
\begin{split}
\int_0^{\frac{2\pi}{\omega}} \frac{\omega dt}{2\pi} f_{p_yt}^{\pm} 
&= \int_0^{\frac{2\pi}{\omega}} \frac{\omega dt}{2\pi}\partial_{p_y} R^m\partial_{t} R^l\epsilon^{mln}b_R^{\pm n}
\\
&= \pm \int_0^{\frac{2\pi}{\omega}} \frac{\omega dt}{2\pi}\frac{1}{2}\frac{p_x\lambda_y\left(\partial_{t} \lambda_x M- \lambda_x \partial_{t} M\right)}{\left(\lambda_x^2p_x^2+\lambda_y^2p_y^2\right)^{\frac{3}{2}}}
\\
&\sim\mp \frac{1}{4}\frac{\omega p_x\left(\lambda_{x0}\Delta_y-\lambda_{y0}\Delta_x\right) \Delta_M\cos\delta}{\left(\lambda_{x0}^2p_x^2+\lambda_{y0}^2p_y^2+M_0^2\right)^{\frac{3}{2}}}
\\
&\pm \frac{3}{4}\frac{\omega p_x\lambda_{y0}^2p_y^2\left(\lambda_{x0}\Delta_y-\lambda_{y0}\Delta_x\right) \Delta_M\cos\delta}{\left(\lambda_{x0}^2p_x^2+\lambda_{y0}^2p_y^2+M_0^2\right)^{\frac{5}{2}}}
.
\end{split}
\label{eq:Berry_curvature3}
\end{align}
Here we decompose $\lambda_x$, $\lambda_y$ and $M$ into nonperturbative static and perturbative time-dependent parts as $\lambda_{x(y)}=\lambda_{x0(y0)}+\Delta_{x(y)} \cos(\omega t)$ and $M=M_0+\Delta_M \sin(\omega t+\delta)$ [$\lambda_{x0(y0)}\gg\Delta_{x(y)}$, and $M_0\gg \Delta_M$].
We consider a monochromatic oscillation with the frequency $\omega$, and evaluate the Berry curvature up to the leading order of $\Delta_{x}$, $\Delta_{y}$, and $\Delta_{M}$.

\subsection{Boltzmann transport theory} 
\label{sec:boltzmann}

We calculate the anomalous currents on the basis of the Boltzmann transport theory with the relaxation time approximation.
In two dimensions, the Boltzmann equation with Berry curvature corrections is given as~\cite{PhysRevB.95.125137}
\bea
&&\frac{\partial n_l}{\partial t} + \dot{\bm x}\cdot\frac{\partial n_l}{\partial \bm x}+ \dot{\bm p}\cdot\frac{\partial n_\sigma}{\partial \bm p}
=-\frac{1}{\tau}\left(n_l-n_{0l}\right), 
\label{eq:Boltzmann} \\
&&\left(1+qBb^\sigma\right) \dot{\bm x}=\tilde{\bm v}+b^\sigma\hat{\bm  z}\times \widetilde{\bm E} ,
\label{eq:sol_x}
\\
&&\left(1+qBb^\sigma\right) \dot{\bm p}=\widetilde{\bm E}+qB\hat{\bm  z}\times\tilde{\bm v},
\eea
where $l=(\sigma,\bm p)$, $\tau$, and $q>0$ denote the electron states (spin and momentum), relaxation time, and elementary charge, respectively.
$n_{0l}=n_0(\tilde{\ve}_l)$ is the equilibrium distribution function with $n_0$ being the Fermi-Dirac distribution function.
$B$ is the magnetic field along the $z$ direction, and $b^\sigma=f^\sigma_{p_xp_y}$ ($\hat{\bm z}$ is the unit vector along the $z$ direction).
We define $\tilde{\bm{v}}$ and $\widetilde{\bm{E}}$ as $\tilde{\bm{v}}\equiv\bm{\nabla}_p\tilde{\ve}_l-\bm e^\sigma$ and $\widetilde{\bm{E}}\equiv -q\bm{E}-\bm{\nabla}_x\tilde{\ve}_l$, 
where $\tilde{\ve}_l=\ve_l-\bm{m}_l\cdot B(\bm x) \hat{\bm z}$ with $\bm{m}_l$ being the orbital magnetic moment, and $\bm e^\sigma=(f^\sigma_{p_xt},f^\sigma_{p_yt})$.
Below we consider static and homogeneous magnetic field and take $\widetilde{\bm{E}}=-q\bm E$. 

By assuming spatially homogeneous distribution, we solve the Boltzmann equation in the $\omega \tau\ll1$ regime. 
The distribution function reads, up to linear order of electric and magnetic fields,
\be
\begin{split}
n_l 
&=n_{0l}+\delta n_{1l}+\delta n_{2l}
\\
&=n_{0l}+\tau\frac{dn_{0l}}{d\ve}q\bm E\cdot\bm v_{0l}+\tau\frac{dn_{0l}}{d\ve}\left(qB \hat{\bm z}\times \bm e^\sigma\right)\cdot\bm v_{0l}, 
\end{split}
\label{eq:Boltzmann_sol}
\ee
where $\bm v_{0l}=\nabla_p\ve_{l}$.
The electric current is given as
\bea
\bm j 
&=&q\sum_l \left(\bm e^\sigma+b^\sigma\hat{\bm  z}\times q\bm E\right) n_{0}(\ve_l)
\nonumber \\
&-&q\sum_l\left(\bm v_{0l}-\bm e^\sigma\right) \delta n_{1l}
-q\sum_l\bm v_{0l} \delta n_{2l} ,
\eea
where we define $\sum_l\equiv\sum_\sigma\int d^2p/(2\pi)^2$, and neglected $\cO(\bm e^2)$ terms, which may suffer from contributions from higher order terms in the kinetic theory.
We note that the Zeeman energy shift does not contribute to anomalous currents in the leading order in $\bm E$ and $\bm B$.
The first term gives two anomalous currents. One is the so called adiabatic charge pumping~\cite{RevModPhys.82.1959} and the other is the intrinsic contribution to the anomalous Hall effect~\cite{Sinitsyn,RevModPhys.82.1539,RevModPhys.82.1959}. 
They are apparently independent of relaxation time and nondissipative.
The second term gives the conventional dissipative current induced by electric fields, and the anomalous one originated from the interplay between $\bm e$ and $\bm E$.
In the Rashba Hamiltonian, the former induces the current parallel to electric fields, but no current transverse to electric fields.
On the other hand, the latter gives no longitudinal  current, and  gives only the transverse current.
The last term is the Hall-type pumping, which is perpendicular to emergent electirc fields $\bm e$, not electric fields $\bm E$ as in $\delta n_{2l}$. The mechanism is similar to the extrinsic anomalous Hall effect because of skew scattering~\cite{Sinitsyn,RevModPhys.82.1539,RevModPhys.82.1959,PhysRevB.68.165311,PhysRevB.75.045315}, in which the change of the distribution function is given as $\delta n_{l}^{\rm tr}\sim \frac{dn_{0l}}{d\ve}(\hat{\bm z}\times \bm E)\cdot \bm v_{0l}$.
We note that these expressions are model-independent.

\subsection{Model calculation}
The intrinsic anomalous Hall effect, that is, the anomalous current originated from emergent magnetic fields has been discussed in details (See e.g., Refs~\cite{Sinitsyn,RevModPhys.82.1539,RevModPhys.82.1959}). 
Then we focus on those induced by emergent electric fields.
We find that the adiabatic and Hall-type charge pumpings vanish because of the symmetry under $p_x\rightarrow -p_x$ and $p_y\rightarrow -p_y$. 
The nonvanishing charge pumping comes from only the interplay between $\bm e$ and $\bm E$, and contributes to the transverse conductivities defined by $j_x=\sigma_{xy}E_y$ and $j_y=\sigma_{yx}E_x$: 
\be
\begin{split}
\sigma_{xy} 
&= q^2\tau\sum_{\sigma=\pm}\int_0^{\frac{2\pi}{\omega}} \frac{\omega dt}{2\pi}\int \frac{d^2p}{(2\pi)^2}f^\sigma_{p_xt} v_{0l}^y \frac{dn_0}{d\ve}(\ve_l)
\\
&=\kappa\sum_{\sigma=\pm}\left(\frac{\tilde{p}_{F\sigma}^3}{R^3_{\bm p_{F\sigma}}}-\frac{3}{4}\frac{\tilde{p}_{F\sigma}^5}{R^5_{\bm p_{F\sigma}}}\right)\left[\frac{\sigma}{m\lambda_{0}^2}+\frac{1}{R_{\bm p_{F\sigma}}}\right] ,
\end{split}
\label{eq:extrinsic_rashba1}
\ee
where $\sigma_{yx}=\sigma_{xy}$, $\kappa=q^2\omega\tau(\Delta_y-\Delta_x) \Delta_M\cos\delta/(16\pi\lambda_0)$, $\tilde{\bm p}=\lambda\bm p$ [$\tilde{p}=|\tilde{\bm p}|$], $R_{\bm p}=R(\bm p)$, $\bm p_{F\sigma}$ are the Fermi momentum, and we consider the zero temperature limit.
$j_x$ and $j_y$ are the average of electric currents over one cycle of the time-periodic external fields.
We here assume that $\lambda_{x0}=\lambda_{y0}=\lambda_0$ to evaluate the momentum integration with the isotropic Fermi surfaces.
In general cases, $\sigma_{xy}\neq \sigma_{yx}$.
We note that for $\mu<|M|$, only the lower band ($\sigma=-$) contributes to the conductivities, and for $\mu<-|M|$, there are contributions from the inside and outside Fermi surfaces of it if $m|M|<\lambda_{0}^2$. 
Eq.~\eqref{eq:extrinsic_rashba1} is originated from the symmetric and longitudinal change of the distribution function $\delta n_{1l}$, 
and the conductivity becomes symmetric, $\sigma_{xy}=\sigma_{yx}$, rather than antisymmetric, $\sigma_{xy}= -\sigma_{yx}$.
Therefore the current (simultaneously) flows in the same direction with the anomalous Hall current, but  contributes to the entropy production ($\bm j\cdot \bm E\neq0$).
If we consider the asymmetric and transverse change e.g.,  due to skew scattering $\delta n_{l}^{\rm tr}\sim \frac{dn_{0l}}{d\ve}(\hat{\bm z}\times \bm E)\cdot \bm v_{0l}$~\cite{Sinitsyn,RevModPhys.82.1539,RevModPhys.82.1959,PhysRevB.68.165311,PhysRevB.75.045315}, the charge pumping $q\bm e^\sigma \delta n_l^{\rm tr}$ now contributes to the longitudinal conductivities $\sigma_{xx}$ and $\sigma_{yy}$ in an anti-symmetric way.

\section{Three dimensional system}
\subsection{Model Hamiltonian}
\label{sec:rashba}

We consider Weyl Hamiltonian as an three-dimensional fermion model to have nonzero Berry curvature. 
The model Hamiltonian is given as~\cite{PhysRevLett.117.216601,2017arXiv170201450I}
\be
  \cH=\chi\sum_a \sigma_a R_a, 
  \label{eq:Hweyl}
\ee
where $\bm R(\bm p)=(v p_x + g D_y,v p_y - g D_x,v_z p_z)^t$, and $\chi=\pm1$ for right- and left-handed Weyl nodes, respectively.
$p_a$ ($a=x,y,z$) are momentum measured from the Weyl node.
We choose the axis of $\bm p$, so that the pair of Weyl nodes related by time-reversal or inversion symmetry are aligned along the $z$ axis. 
$v$ and $v_z$ are velocities of Weyl electrons along the $x$ ($y$) and $z$ directions, which in general take different values due to the uniaxial anisotropy about the $z$ axis.
$D_a$ ($a=x,y$) are AC electric field along the $x$ and $y$ directions.
As the coupling to the external electric field, we consider an orbital coupling with the coupling strength $g$, which is allowed by symmetry when the Weyl node is put away from high-symmetry points, such as the $\Gamma$ point~\cite{PhysRevLett.117.216601,2017arXiv170201450I}.

We consider the incident light along the $z$ direction $\bm D=(D\cos(\omega t),D\sin(\omega t+\delta),0)$,
where $D$, $\omega$, and $\delta$ are, respectively, the square root of the intensity,  the frequency, and the phase shift of light. 
$\bm D$ describes a circularly polarized light when $\delta=0,\pi$, and a linearly polarized light when $\delta=\pi/2, 3\pi/2$. 
The Berry curvature in $\bm R$ space is the same as that of the Rashba Hamiltonian.
We can calculate the Berry curvature in physical space $\xi_\mu=(t,p_x,p_y,p_z)$ in the same way with the Rashba Hamiltonian by using the pullback.

\subsection{Boltzmann transport theory} 
\label{sec:boltzmann}

In three dimensions, the Boltzmann equation with Berry curvature corrections is now given by Eq.~\eqref{eq:Boltzmann} 
and the following classical equation of motion~\cite{RevModPhys.82.1959,PhysRevB.95.125137}: 
\begin{align}
&\left(1+q\bm B\cdot \bm b^\sigma\right) \dot{\bm x}=\tilde{\bm v}-\bm b^\sigma\times q\widetilde{\bm E}+\left(\tilde{\bm v}\cdot\bm b^\sigma\right)q\bm B ,
\label{eq:sol_x}
\\
&\left(1+q\bm B\cdot \bm b^\sigma\right) \dot{\bm p}=-q\widetilde{\bm E}+q\bm B\times\tilde{\bm v}-\left(q\widetilde{\bm E}\cdot q\bm B\right)\bm b^\sigma,
\end{align}
where $\tilde{\bm{v}}=\bm{\nabla}_p\ve_l-\bm e^\sigma$, $\bm e^\sigma=(f^\sigma_{p_xt},f^\sigma_{p_yt},f^\sigma_{p_zt})$, and $\widetilde{\bm E}=\bm E+\bm D$. 
$\bm E$ and $\bm D$ are DC and AC external electric fields, respectively~\cite{2017arXiv170201450I}. 
We neglected the Zeeman energy shift, which does not modify the following expressions.

The change of spatially homogeneous distribution function is given as, up to linear order of electric and magnetic fields,
\be
\begin{split}
n_l 
&=n_{0l}+\delta n_{1l}+\delta n_{2l}
\\
&=n_{0l}+\tau\frac{dn_{0l}}{d\ve}q\widetilde{\bm E}\cdot\bm v_{0l}+\tau\frac{dn_{0l}}{d\ve}\left(q\bm B \times \bm e^\sigma\right)\cdot\bm v_{0l}, 
\end{split}
\label{eq:Boltzmann_sol}
\ee
where $\bm v_{0l}=\nabla_p\ve_{l}$.
The electric current is given as
\bea
\bm j 
&=&q\sum_l \left(\bm e^\sigma+\bm b^\sigma\times q\widetilde{\bm E}-\left(\tilde{\bm v}\cdot\bm b^\sigma\right)q\bm B\right) n_{0}(\ve_l)
\nonumber \\
&-&q\sum_l\left(\bm v_{0l}-\bm e^\sigma\right) \delta n_{1l}
-q\sum_l\bm v_{0l} \delta n_{2l} ,
\eea
where we define $\sum_l\equiv\sum_\sigma\int d^3p/(2\pi)^3$, and neglected $\cO(\bm e^2)$ terms.
The first term gives the adiabatic charge pumping~\cite{RevModPhys.82.1959}, 
the intrinsic anomalous Hall effect~\cite{Sinitsyn,RevModPhys.82.1539,RevModPhys.82.1959}, 
and the chiral magnetic effect~\cite{PhysRevLett.109.162001,PhysRevLett.109.181602,PhysRevLett.110.262301}. 
They are obtained from the equilibrium distribution function, and nondissipative.
The second term gives the conventional dissipative current induced by electric fields, and the dissipative charge pumping originated from the interplay between $\bm e$ and $\widetilde{\bm E}$.
The last term is the Hall-type charge pumping transverse to the emergent electric fields.
As shown below, it leads to a new type of DC Hall current.
We note that these expressions are model-independent.

\subsection{Model calculation}
The anomalous currents originated from emergent magnetic fields have been discussed in details.
Then we focus on the anomalous currents induced by emergent electric fields.
We find that the adiabatic charge pumping, and the dissipative charge pumping originated from the interplay between $\bm e$ and $\widetilde{\bm E}$ vanish in the linear Weyl model.
The nonvanishing charge pumping comes from only the interplay between $\bm e$ and $\bm B$, and is given as, in the leading order of $D$, 
\bea
j_x&=&\tilde{\sigma}_{xy} B_y ,\\
\tilde{\sigma}_{xy} &=& \mp q^2\tau\int_0^{\frac{2\pi}{\omega}} \frac{\omega dt}{2\pi}\int \frac{d^3p}{(2\pi)^3}f^+_{p_zt} \left(v_{0l}^x\right)^2 \frac{dn_0}{d\ve}(\ve_l)
\nonumber \\
&&\pm q^2\tau\int_0^{\frac{2\pi}{\omega}} \frac{\omega dt}{2\pi}\int \frac{d^3p}{(2\pi)^3}f^+_{p_xt} v_{0l}^xv_{0l}^z \frac{dn_0}{d\ve}(\ve_l)
\nonumber \\
&=& \frac{\chi q^2\omega\tau g^2D^2\cos\delta}{24\pi^2\mu} ,
\label{eq:extrinsic_weyl}
\eea
where $\tilde{\sigma}_{yx}=-\tilde{\sigma}_{xy}$, $j_z=0$, $\mu$ is the chemical potential measured from the Weyl node energy, 
and we consider the zero temperature limit. 
The expression is the same for positive ($\mu>0$) and negative doping ($\mu<0$).
The conductivity takes the maximum value in circularly polarized light, while vanishes in linearly polarized light because of the polarization-dependence of the emergent electric fields~\cite{PhysRevLett.117.216601,2017arXiv170201450I}.
This is the Hall-type photocurrent, which flows in the direction transverse to DC magnetic fields.
Since the electron's orbital coupling to electric fields, and the chemical potential measured from the Weyl node can differ between right- and left-handed Weyl nodes in the absence of the inversion symmetry~\cite{PhysRevLett.117.216601,2017arXiv170201450I},
the Hall photocurrent does not cancel between Weyl nodes, and can contribute to transport phenomena in the inversion symmetry breaking Weyl semimetals such as TaAs or SrSi$_2$~\cite{Huang,PhysRevX.5.011029,PhysRevX.5.031013,Xu613,Huang02022016}.

In the Weyl semimetals without tilting or nonlinear term, the adiabatic charge pumping vanishes.
Thus the Hall photocurrent is the first nontrivial effect induced by emergent electric fields.
Even if we take them into account, the photocurrent from the adiabatic charge pumping is parallel to the Weyl node separation vector ($\hat{\bm z}$ in our notation)~\cite{PhysRevLett.117.216601,2017arXiv170201450I}. 
Therefore these effects are perpendicular to each other and in general distinguishable. 

\section{Summary}  

We have studied the dissipative currents originated from the interplay between emergent electric fields and electric/magnetic fields in two- and three-dimensional models on the basis of the Boltzmann transport theory.
In two dimensions, by studying the Rashba Hamiltonian, 
we show that the interplay between emergent electric fields and electric fields leads to the anomalous current transverse to electric fields.
The dissipative charge pumping is not anti-symmetric and contributes to the entropy production.
In three dimensions, by studying the Weyl Hamiltonian, we show that the interplay between emergent electric fields and magnetic fields leads to the dissipative Hall current in the absence of DC electric fields, which is now transverse to DC magnetic fields.
The Hall photocurrent will be relevant in the inversion symmetry breaking Weyl semimetals such as TaAs or SrSi$_2$~\cite{Huang,PhysRevX.5.011029,PhysRevX.5.031013,Xu613,Huang02022016}.

There are several generalizations of our work.
One direction is to consider spatially inhomogeneous change of the distribution function and calculate the Hall conductivity at finite-momentum~\cite{PhysRevLett.108.066805,PhysRevLett.118.226601}. 
It is also interesting to study thermal currents and viscosities.
Another direction is to study higher order terms in electromagnetic fields, that is, nonlinear transport phenomena~\cite{Morimotoe1501524,PhysRevB.94.235123,PhysRevB.94.245121} with taking the effect of emergent electric fields into account.

We can perform the same analysis in the presence of the nonabelian Berry curvature on the basis of the nonabelian version of the kinetic theory~\cite{PhysRevB.72.085110,Shindou2005399,0953-8984-20-19-193202,Hayata:2017ihy}.
There may arise the dissipative transverse spin pumping or the dissipative $B$-induced Hall-type spin pumping, which is originated from the interplay between emergent electric fields and electric or magnetic fields e.g., in graphene or Dirac semimetals.
 
%%\\
\begin{acknowledgements}
This work was supported by JSPS Grant-in-Aid for Scientific Research (No: JP16J02240).
\end{acknowledgements}

%------------------------------------------------------------

\bibliography{./rashba}

\begin{thebibliography}{56}%
\makeatletter
\providecommand \@ifxundefined [1]{%
 \@ifx{#1\undefined}
}%
\providecommand \@ifnum [1]{%
 \ifnum #1\expandafter \@firstoftwo
 \else \expandafter \@secondoftwo
 \fi
}%
\providecommand \@ifx [1]{%
 \ifx #1\expandafter \@firstoftwo
 \else \expandafter \@secondoftwo
 \fi
}%
\providecommand \natexlab [1]{#1}%
\providecommand \enquote  [1]{``#1''}%
\providecommand \bibnamefont  [1]{#1}%
\providecommand \bibfnamefont [1]{#1}%
\providecommand \citenamefont [1]{#1}%
\providecommand \href@noop [0]{\@secondoftwo}%
\providecommand \href [0]{\begingroup \@sanitize@url \@href}%
\providecommand \@href[1]{\@@startlink{#1}\@@href}%
\providecommand \@@href[1]{\endgroup#1\@@endlink}%
\providecommand \@sanitize@url [0]{\catcode `\\12\catcode `\$12\catcode
  `\&12\catcode `\#12\catcode `\^12\catcode `\_12\catcode `\%12\relax}%
\providecommand \@@startlink[1]{}%
\providecommand \@@endlink[0]{}%
\providecommand \url  [0]{\begingroup\@sanitize@url \@url }%
\providecommand \@url [1]{\endgroup\@href {#1}{\urlprefix }}%
\providecommand \urlprefix  [0]{URL }%
\providecommand \Eprint [0]{\href }%
\providecommand \doibase [0]{http://dx.doi.org/}%
\providecommand \selectlanguage [0]{\@gobble}%
\providecommand \bibinfo  [0]{\@secondoftwo}%
\providecommand \bibfield  [0]{\@secondoftwo}%
\providecommand \translation [1]{[#1]}%
\providecommand \BibitemOpen [0]{}%
\providecommand \bibitemStop [0]{}%
\providecommand \bibitemNoStop [0]{.\EOS\space}%
\providecommand \EOS [0]{\spacefactor3000\relax}%
\providecommand \BibitemShut  [1]{\csname bibitem#1\endcsname}%
\let\auto@bib@innerbib\@empty
%</preamble>
\bibitem [{\citenamefont {Klitzing}\ \emph {et~al.}(1980)\citenamefont
  {Klitzing}, \citenamefont {Dorda},\ and\ \citenamefont
  {Pepper}}]{PhysRevLett.45.494}%
  \BibitemOpen
  \bibfield  {author} {\bibinfo {author} {\bibfnamefont {K.~v.}\ \bibnamefont
  {Klitzing}}, \bibinfo {author} {\bibfnamefont {G.}~\bibnamefont {Dorda}}, \
  and\ \bibinfo {author} {\bibfnamefont {M.}~\bibnamefont {Pepper}},\ }\href
  {http://link.aps.org/doi/10.1103/PhysRevLett.45.494} {\bibfield  {journal}
  {\bibinfo  {journal} {Phys. Rev. Lett.}\ }\textbf {\bibinfo {volume} {45}},\
  \bibinfo {pages} {494} (\bibinfo {year} {1980})}\BibitemShut {NoStop}%
\bibitem [{\citenamefont {Laughlin}(1981)}]{PhysRevB.23.5632}%
  \BibitemOpen
  \bibfield  {author} {\bibinfo {author} {\bibfnamefont {R.~B.}\ \bibnamefont
  {Laughlin}},\ }\href {http://link.aps.org/doi/10.1103/PhysRevB.23.5632}
  {\bibfield  {journal} {\bibinfo  {journal} {Phys. Rev. B}\ }\textbf {\bibinfo
  {volume} {23}},\ \bibinfo {pages} {5632} (\bibinfo {year}
  {1981})}\BibitemShut {NoStop}%
\bibitem [{\citenamefont {Thouless}\ \emph {et~al.}(1982)\citenamefont
  {Thouless}, \citenamefont {Kohmoto}, \citenamefont {Nightingale},\ and\
  \citenamefont {den Nijs}}]{PhysRevLett.49.405}%
  \BibitemOpen
  \bibfield  {author} {\bibinfo {author} {\bibfnamefont {D.~J.}\ \bibnamefont
  {Thouless}}, \bibinfo {author} {\bibfnamefont {M.}~\bibnamefont {Kohmoto}},
  \bibinfo {author} {\bibfnamefont {M.~P.}\ \bibnamefont {Nightingale}}, \ and\
  \bibinfo {author} {\bibfnamefont {M.}~\bibnamefont {den Nijs}},\ }\href
  {http://link.aps.org/doi/10.1103/PhysRevLett.49.405} {\bibfield  {journal}
  {\bibinfo  {journal} {Phys. Rev. Lett.}\ }\textbf {\bibinfo {volume} {49}},\
  \bibinfo {pages} {405} (\bibinfo {year} {1982})}\BibitemShut {NoStop}%
\bibitem [{\citenamefont {Karplus}\ and\ \citenamefont
  {Luttinger}(1954)}]{PhysRev.95.1154}%
  \BibitemOpen
  \bibfield  {author} {\bibinfo {author} {\bibfnamefont {R.}~\bibnamefont
  {Karplus}}\ and\ \bibinfo {author} {\bibfnamefont {J.~M.}\ \bibnamefont
  {Luttinger}},\ }\href {http://link.aps.org/doi/10.1103/PhysRev.95.1154}
  {\bibfield  {journal} {\bibinfo  {journal} {Phys. Rev.}\ }\textbf {\bibinfo
  {volume} {95}},\ \bibinfo {pages} {1154} (\bibinfo {year}
  {1954})}\BibitemShut {NoStop}%
\bibitem [{\citenamefont {Jungwirth}\ \emph {et~al.}(2002)\citenamefont
  {Jungwirth}, \citenamefont {Niu},\ and\ \citenamefont
  {MacDonald}}]{PhysRevLett.88.207208}%
  \BibitemOpen
  \bibfield  {author} {\bibinfo {author} {\bibfnamefont {T.}~\bibnamefont
  {Jungwirth}}, \bibinfo {author} {\bibfnamefont {Q.}~\bibnamefont {Niu}}, \
  and\ \bibinfo {author} {\bibfnamefont {A.~H.}\ \bibnamefont {MacDonald}},\
  }\href {http://link.aps.org/doi/10.1103/PhysRevLett.88.207208} {\bibfield
  {journal} {\bibinfo  {journal} {Phys. Rev. Lett.}\ }\textbf {\bibinfo
  {volume} {88}},\ \bibinfo {pages} {207208} (\bibinfo {year}
  {2002})}\BibitemShut {NoStop}%
\bibitem [{\citenamefont {Haldane}(2004)}]{PhysRevLett.93.206602}%
  \BibitemOpen
  \bibfield  {author} {\bibinfo {author} {\bibfnamefont {F.~D.~M.}\
  \bibnamefont {Haldane}},\ }\href
  {http://link.aps.org/doi/10.1103/PhysRevLett.93.206602} {\bibfield  {journal}
  {\bibinfo  {journal} {Phys. Rev. Lett.}\ }\textbf {\bibinfo {volume} {93}},\
  \bibinfo {pages} {206602} (\bibinfo {year} {2004})}\BibitemShut {NoStop}%
\bibitem [{\citenamefont {Sinitsyn}(2008)}]{Sinitsyn}%
  \BibitemOpen
  \bibfield  {author} {\bibinfo {author} {\bibfnamefont {N.~A.}\ \bibnamefont
  {Sinitsyn}},\ }\href {http://stacks.iop.org/0953-8984/20/i=2/a=023201}
  {\bibfield  {journal} {\bibinfo  {journal} {Journal of Physics: Condensed
  Matter}\ }\textbf {\bibinfo {volume} {20}},\ \bibinfo {pages} {023201}
  (\bibinfo {year} {2008})}\BibitemShut {NoStop}%
\bibitem [{\citenamefont {Nagaosa}\ \emph {et~al.}(2010)\citenamefont
  {Nagaosa}, \citenamefont {Sinova}, \citenamefont {Onoda}, \citenamefont
  {MacDonald},\ and\ \citenamefont {Ong}}]{RevModPhys.82.1539}%
  \BibitemOpen
  \bibfield  {author} {\bibinfo {author} {\bibfnamefont {N.}~\bibnamefont
  {Nagaosa}}, \bibinfo {author} {\bibfnamefont {J.}~\bibnamefont {Sinova}},
  \bibinfo {author} {\bibfnamefont {S.}~\bibnamefont {Onoda}}, \bibinfo
  {author} {\bibfnamefont {A.~H.}\ \bibnamefont {MacDonald}}, \ and\ \bibinfo
  {author} {\bibfnamefont {N.~P.}\ \bibnamefont {Ong}},\ }\href
  {http://link.aps.org/doi/10.1103/RevModPhys.82.1539} {\bibfield  {journal}
  {\bibinfo  {journal} {Rev. Mod. Phys.}\ }\textbf {\bibinfo {volume} {82}},\
  \bibinfo {pages} {1539} (\bibinfo {year} {2010})}\BibitemShut {NoStop}%
\bibitem [{\citenamefont {Xiao}\ \emph {et~al.}(2010)\citenamefont {Xiao},
  \citenamefont {Chang},\ and\ \citenamefont {Niu}}]{RevModPhys.82.1959}%
  \BibitemOpen
  \bibfield  {author} {\bibinfo {author} {\bibfnamefont {D.}~\bibnamefont
  {Xiao}}, \bibinfo {author} {\bibfnamefont {M.-C.}\ \bibnamefont {Chang}}, \
  and\ \bibinfo {author} {\bibfnamefont {Q.}~\bibnamefont {Niu}},\ }\href
  {http://link.aps.org/doi/10.1103/RevModPhys.82.1959} {\bibfield  {journal}
  {\bibinfo  {journal} {Rev. Mod. Phys.}\ }\textbf {\bibinfo {volume} {82}},\
  \bibinfo {pages} {1959} (\bibinfo {year} {2010})}\BibitemShut {NoStop}%
\bibitem [{\citenamefont {Yu}\ \emph {et~al.}(2010)\citenamefont {Yu},
  \citenamefont {Zhang}, \citenamefont {Zhang}, \citenamefont {Zhang},
  \citenamefont {Dai},\ and\ \citenamefont {Fang}}]{Yu61}%
  \BibitemOpen
  \bibfield  {author} {\bibinfo {author} {\bibfnamefont {R.}~\bibnamefont
  {Yu}}, \bibinfo {author} {\bibfnamefont {W.}~\bibnamefont {Zhang}}, \bibinfo
  {author} {\bibfnamefont {H.-J.}\ \bibnamefont {Zhang}}, \bibinfo {author}
  {\bibfnamefont {S.-C.}\ \bibnamefont {Zhang}}, \bibinfo {author}
  {\bibfnamefont {X.}~\bibnamefont {Dai}}, \ and\ \bibinfo {author}
  {\bibfnamefont {Z.}~\bibnamefont {Fang}},\ }\href
  {http://science.sciencemag.org/content/329/5987/61} {\bibfield  {journal}
  {\bibinfo  {journal} {Science}\ }\textbf {\bibinfo {volume} {329}},\ \bibinfo
  {pages} {61} (\bibinfo {year} {2010})}\BibitemShut {NoStop}%
\bibitem [{\citenamefont {Chang}\ \emph {et~al.}(2013)\citenamefont {Chang},
  \citenamefont {Zhang}, \citenamefont {Feng}, \citenamefont {Shen},
  \citenamefont {Zhang}, \citenamefont {Guo}, \citenamefont {Li}, \citenamefont
  {Ou}, \citenamefont {Wei}, \citenamefont {Wang}, \citenamefont {Ji},
  \citenamefont {Feng}, \citenamefont {Ji}, \citenamefont {Chen}, \citenamefont
  {Jia}, \citenamefont {Dai}, \citenamefont {Fang}, \citenamefont {Zhang},
  \citenamefont {He}, \citenamefont {Wang}, \citenamefont {Lu}, \citenamefont
  {Ma},\ and\ \citenamefont {Xue}}]{Chang167}%
  \BibitemOpen
  \bibfield  {author} {\bibinfo {author} {\bibfnamefont {C.-Z.}\ \bibnamefont
  {Chang}}, \bibinfo {author} {\bibfnamefont {J.}~\bibnamefont {Zhang}},
  \bibinfo {author} {\bibfnamefont {X.}~\bibnamefont {Feng}}, \bibinfo {author}
  {\bibfnamefont {J.}~\bibnamefont {Shen}}, \bibinfo {author} {\bibfnamefont
  {Z.}~\bibnamefont {Zhang}}, \bibinfo {author} {\bibfnamefont
  {M.}~\bibnamefont {Guo}}, \bibinfo {author} {\bibfnamefont {K.}~\bibnamefont
  {Li}}, \bibinfo {author} {\bibfnamefont {Y.}~\bibnamefont {Ou}}, \bibinfo
  {author} {\bibfnamefont {P.}~\bibnamefont {Wei}}, \bibinfo {author}
  {\bibfnamefont {L.-L.}\ \bibnamefont {Wang}}, \bibinfo {author}
  {\bibfnamefont {Z.-Q.}\ \bibnamefont {Ji}}, \bibinfo {author} {\bibfnamefont
  {Y.}~\bibnamefont {Feng}}, \bibinfo {author} {\bibfnamefont {S.}~\bibnamefont
  {Ji}}, \bibinfo {author} {\bibfnamefont {X.}~\bibnamefont {Chen}}, \bibinfo
  {author} {\bibfnamefont {J.}~\bibnamefont {Jia}}, \bibinfo {author}
  {\bibfnamefont {X.}~\bibnamefont {Dai}}, \bibinfo {author} {\bibfnamefont
  {Z.}~\bibnamefont {Fang}}, \bibinfo {author} {\bibfnamefont {S.-C.}\
  \bibnamefont {Zhang}}, \bibinfo {author} {\bibfnamefont {K.}~\bibnamefont
  {He}}, \bibinfo {author} {\bibfnamefont {Y.}~\bibnamefont {Wang}}, \bibinfo
  {author} {\bibfnamefont {L.}~\bibnamefont {Lu}}, \bibinfo {author}
  {\bibfnamefont {X.-C.}\ \bibnamefont {Ma}}, \ and\ \bibinfo {author}
  {\bibfnamefont {Q.-K.}\ \bibnamefont {Xue}},\ }\href
  {http://science.sciencemag.org/content/340/6129/167} {\bibfield  {journal}
  {\bibinfo  {journal} {Science}\ }\textbf {\bibinfo {volume} {340}},\ \bibinfo
  {pages} {167} (\bibinfo {year} {2013})}\BibitemShut {NoStop}%
\bibitem [{\citenamefont {Hirsch}(1999)}]{PhysRevLett.83.1834}%
  \BibitemOpen
  \bibfield  {author} {\bibinfo {author} {\bibfnamefont {J.~E.}\ \bibnamefont
  {Hirsch}},\ }\href {https://link.aps.org/doi/10.1103/PhysRevLett.83.1834}
  {\bibfield  {journal} {\bibinfo  {journal} {Phys. Rev. Lett.}\ }\textbf
  {\bibinfo {volume} {83}},\ \bibinfo {pages} {1834} (\bibinfo {year}
  {1999})}\BibitemShut {NoStop}%
\bibitem [{\citenamefont {Sinova}\ \emph {et~al.}(2004)\citenamefont {Sinova},
  \citenamefont {Culcer}, \citenamefont {Niu}, \citenamefont {Sinitsyn},
  \citenamefont {Jungwirth},\ and\ \citenamefont
  {MacDonald}}]{PhysRevLett.92.126603}%
  \BibitemOpen
  \bibfield  {author} {\bibinfo {author} {\bibfnamefont {J.}~\bibnamefont
  {Sinova}}, \bibinfo {author} {\bibfnamefont {D.}~\bibnamefont {Culcer}},
  \bibinfo {author} {\bibfnamefont {Q.}~\bibnamefont {Niu}}, \bibinfo {author}
  {\bibfnamefont {N.~A.}\ \bibnamefont {Sinitsyn}}, \bibinfo {author}
  {\bibfnamefont {T.}~\bibnamefont {Jungwirth}}, \ and\ \bibinfo {author}
  {\bibfnamefont {A.~H.}\ \bibnamefont {MacDonald}},\ }\href
  {https://link.aps.org/doi/10.1103/PhysRevLett.92.126603} {\bibfield
  {journal} {\bibinfo  {journal} {Phys. Rev. Lett.}\ }\textbf {\bibinfo
  {volume} {92}},\ \bibinfo {pages} {126603} (\bibinfo {year}
  {2004})}\BibitemShut {NoStop}%
\bibitem [{\citenamefont {Kato}\ \emph {et~al.}(2004)\citenamefont {Kato},
  \citenamefont {Myers}, \citenamefont {Gossard},\ and\ \citenamefont
  {Awschalom}}]{Kato1910}%
  \BibitemOpen
  \bibfield  {author} {\bibinfo {author} {\bibfnamefont {Y.~K.}\ \bibnamefont
  {Kato}}, \bibinfo {author} {\bibfnamefont {R.~C.}\ \bibnamefont {Myers}},
  \bibinfo {author} {\bibfnamefont {A.~C.}\ \bibnamefont {Gossard}}, \ and\
  \bibinfo {author} {\bibfnamefont {D.~D.}\ \bibnamefont {Awschalom}},\ }\href
  {http://science.sciencemag.org/content/306/5703/1910} {\bibfield  {journal}
  {\bibinfo  {journal} {Science}\ }\textbf {\bibinfo {volume} {306}},\ \bibinfo
  {pages} {1910} (\bibinfo {year} {2004})}\BibitemShut {NoStop}%
\bibitem [{\citenamefont {Wunderlich}\ \emph {et~al.}(2005)\citenamefont
  {Wunderlich}, \citenamefont {Kaestner}, \citenamefont {Sinova},\ and\
  \citenamefont {Jungwirth}}]{PhysRevLett.94.047204}%
  \BibitemOpen
  \bibfield  {author} {\bibinfo {author} {\bibfnamefont {J.}~\bibnamefont
  {Wunderlich}}, \bibinfo {author} {\bibfnamefont {B.}~\bibnamefont
  {Kaestner}}, \bibinfo {author} {\bibfnamefont {J.}~\bibnamefont {Sinova}}, \
  and\ \bibinfo {author} {\bibfnamefont {T.}~\bibnamefont {Jungwirth}},\ }\href
  {https://link.aps.org/doi/10.1103/PhysRevLett.94.047204} {\bibfield
  {journal} {\bibinfo  {journal} {Phys. Rev. Lett.}\ }\textbf {\bibinfo
  {volume} {94}},\ \bibinfo {pages} {047204} (\bibinfo {year}
  {2005})}\BibitemShut {NoStop}%
\bibitem [{\citenamefont {Valenzuela}\ and\ \citenamefont
  {Tinkham}(2006)}]{2006Natur.442..176V}%
  \BibitemOpen
  \bibfield  {author} {\bibinfo {author} {\bibfnamefont {S.~O.}\ \bibnamefont
  {Valenzuela}}\ and\ \bibinfo {author} {\bibfnamefont {M.}~\bibnamefont
  {Tinkham}},\ }\href {http://dx.doi.org/10.1038/nature04937} {\bibfield
  {journal} {\bibinfo  {journal} {Nature}\ }\textbf {\bibinfo {volume} {442}},\
  \bibinfo {pages} {176} (\bibinfo {year} {2006})}\BibitemShut {NoStop}%
\bibitem [{\citenamefont {Kane}\ and\ \citenamefont
  {Mele}(2005)}]{PhysRevLett.95.226801}%
  \BibitemOpen
  \bibfield  {author} {\bibinfo {author} {\bibfnamefont {C.~L.}\ \bibnamefont
  {Kane}}\ and\ \bibinfo {author} {\bibfnamefont {E.~J.}\ \bibnamefont
  {Mele}},\ }\href {https://link.aps.org/doi/10.1103/PhysRevLett.95.226801}
  {\bibfield  {journal} {\bibinfo  {journal} {Phys. Rev. Lett.}\ }\textbf
  {\bibinfo {volume} {95}},\ \bibinfo {pages} {226801} (\bibinfo {year}
  {2005})}\BibitemShut {NoStop}%
\bibitem [{\citenamefont {Bernevig}\ \emph {et~al.}(2006)\citenamefont
  {Bernevig}, \citenamefont {Hughes},\ and\ \citenamefont
  {Zhang}}]{Bernevig1757}%
  \BibitemOpen
  \bibfield  {author} {\bibinfo {author} {\bibfnamefont {B.~A.}\ \bibnamefont
  {Bernevig}}, \bibinfo {author} {\bibfnamefont {T.~L.}\ \bibnamefont
  {Hughes}}, \ and\ \bibinfo {author} {\bibfnamefont {S.-C.}\ \bibnamefont
  {Zhang}},\ }\href {http://science.sciencemag.org/content/314/5806/1757}
  {\bibfield  {journal} {\bibinfo  {journal} {Science}\ }\textbf {\bibinfo
  {volume} {314}},\ \bibinfo {pages} {1757} (\bibinfo {year}
  {2006})}\BibitemShut {NoStop}%
\bibitem [{\citenamefont {Nielsen}\ and\ \citenamefont
  {Ninomiya}(1983)}]{NIELSEN1983389}%
  \BibitemOpen
  \bibfield  {author} {\bibinfo {author} {\bibfnamefont {H.}~\bibnamefont
  {Nielsen}}\ and\ \bibinfo {author} {\bibfnamefont {M.}~\bibnamefont
  {Ninomiya}},\ }\href
  {http://www.sciencedirect.com/science/article/pii/0370269383915290}
  {\bibfield  {journal} {\bibinfo  {journal} {Physics Letters B}\ }\textbf
  {\bibinfo {volume} {130}},\ \bibinfo {pages} {389 } (\bibinfo {year}
  {1983})}\BibitemShut {NoStop}%
\bibitem [{\citenamefont {Fukushima}\ \emph {et~al.}(2008)\citenamefont
  {Fukushima}, \citenamefont {Kharzeev},\ and\ \citenamefont
  {Warringa}}]{PhysRevD.78.074033}%
  \BibitemOpen
  \bibfield  {author} {\bibinfo {author} {\bibfnamefont {K.}~\bibnamefont
  {Fukushima}}, \bibinfo {author} {\bibfnamefont {D.~E.}\ \bibnamefont
  {Kharzeev}}, \ and\ \bibinfo {author} {\bibfnamefont {H.~J.}\ \bibnamefont
  {Warringa}},\ }\href {http://link.aps.org/doi/10.1103/PhysRevD.78.074033}
  {\bibfield  {journal} {\bibinfo  {journal} {Phys. Rev. D}\ }\textbf {\bibinfo
  {volume} {78}},\ \bibinfo {pages} {074033} (\bibinfo {year}
  {2008})}\BibitemShut {NoStop}%
\bibitem [{\citenamefont {Son}\ and\ \citenamefont
  {Spivak}(2013)}]{PhysRevB.88.104412}%
  \BibitemOpen
  \bibfield  {author} {\bibinfo {author} {\bibfnamefont {D.~T.}\ \bibnamefont
  {Son}}\ and\ \bibinfo {author} {\bibfnamefont {B.~Z.}\ \bibnamefont
  {Spivak}},\ }\href {http://link.aps.org/doi/10.1103/PhysRevB.88.104412}
  {\bibfield  {journal} {\bibinfo  {journal} {Phys. Rev. B}\ }\textbf {\bibinfo
  {volume} {88}},\ \bibinfo {pages} {104412} (\bibinfo {year}
  {2013})}\BibitemShut {NoStop}%
\bibitem [{\citenamefont {Huang}\ \emph
  {et~al.}(2015{\natexlab{a}})\citenamefont {Huang}, \citenamefont {Zhao},
  \citenamefont {Long}, \citenamefont {Wang}, \citenamefont {Chen},
  \citenamefont {Yang}, \citenamefont {Liang}, \citenamefont {Xue},
  \citenamefont {Weng}, \citenamefont {Fang}, \citenamefont {Dai},\ and\
  \citenamefont {Chen}}]{PhysRevX.5.031023}%
  \BibitemOpen
  \bibfield  {author} {\bibinfo {author} {\bibfnamefont {X.}~\bibnamefont
  {Huang}}, \bibinfo {author} {\bibfnamefont {L.}~\bibnamefont {Zhao}},
  \bibinfo {author} {\bibfnamefont {Y.}~\bibnamefont {Long}}, \bibinfo {author}
  {\bibfnamefont {P.}~\bibnamefont {Wang}}, \bibinfo {author} {\bibfnamefont
  {D.}~\bibnamefont {Chen}}, \bibinfo {author} {\bibfnamefont {Z.}~\bibnamefont
  {Yang}}, \bibinfo {author} {\bibfnamefont {H.}~\bibnamefont {Liang}},
  \bibinfo {author} {\bibfnamefont {M.}~\bibnamefont {Xue}}, \bibinfo {author}
  {\bibfnamefont {H.}~\bibnamefont {Weng}}, \bibinfo {author} {\bibfnamefont
  {Z.}~\bibnamefont {Fang}}, \bibinfo {author} {\bibfnamefont {X.}~\bibnamefont
  {Dai}}, \ and\ \bibinfo {author} {\bibfnamefont {G.}~\bibnamefont {Chen}},\
  }\href {https://link.aps.org/doi/10.1103/PhysRevX.5.031023} {\bibfield
  {journal} {\bibinfo  {journal} {Phys. Rev. X}\ }\textbf {\bibinfo {volume}
  {5}},\ \bibinfo {pages} {031023} (\bibinfo {year}
  {2015}{\natexlab{a}})}\BibitemShut {NoStop}%
\bibitem [{\citenamefont {Berry}(1984)}]{Berry45}%
  \BibitemOpen
  \bibfield  {author} {\bibinfo {author} {\bibfnamefont {M.~V.}\ \bibnamefont
  {Berry}},\ }\href
  {http://rspa.royalsocietypublishing.org/content/392/1802/45} {\bibfield
  {journal} {\bibinfo  {journal} {Proceedings of the Royal Society of London A:
  Mathematical, Physical and Engineering Sciences}\ }\textbf {\bibinfo {volume}
  {392}},\ \bibinfo {pages} {45} (\bibinfo {year} {1984})}\BibitemShut
  {NoStop}%
\bibitem [{\citenamefont {Sundaram}\ and\ \citenamefont
  {Niu}(1999)}]{PhysRevB.59.14915}%
  \BibitemOpen
  \bibfield  {author} {\bibinfo {author} {\bibfnamefont {G.}~\bibnamefont
  {Sundaram}}\ and\ \bibinfo {author} {\bibfnamefont {Q.}~\bibnamefont {Niu}},\
  }\href {http://link.aps.org/doi/10.1103/PhysRevB.59.14915} {\bibfield
  {journal} {\bibinfo  {journal} {Phys. Rev. B}\ }\textbf {\bibinfo {volume}
  {59}},\ \bibinfo {pages} {14915} (\bibinfo {year} {1999})}\BibitemShut
  {NoStop}%
\bibitem [{\citenamefont {Thouless}(1983)}]{PhysRevB.27.6083}%
  \BibitemOpen
  \bibfield  {author} {\bibinfo {author} {\bibfnamefont {D.~J.}\ \bibnamefont
  {Thouless}},\ }\href {http://link.aps.org/doi/10.1103/PhysRevB.27.6083}
  {\bibfield  {journal} {\bibinfo  {journal} {Phys. Rev. B}\ }\textbf {\bibinfo
  {volume} {27}},\ \bibinfo {pages} {6083} (\bibinfo {year}
  {1983})}\BibitemShut {NoStop}%
\bibitem [{\citenamefont {King-Smith}\ and\ \citenamefont
  {Vanderbilt}(1993)}]{PhysRevB.47.1651}%
  \BibitemOpen
  \bibfield  {author} {\bibinfo {author} {\bibfnamefont {R.~D.}\ \bibnamefont
  {King-Smith}}\ and\ \bibinfo {author} {\bibfnamefont {D.}~\bibnamefont
  {Vanderbilt}},\ }\href {http://link.aps.org/doi/10.1103/PhysRevB.47.1651}
  {\bibfield  {journal} {\bibinfo  {journal} {Phys. Rev. B}\ }\textbf {\bibinfo
  {volume} {47}},\ \bibinfo {pages} {1651} (\bibinfo {year}
  {1993})}\BibitemShut {NoStop}%
\bibitem [{\citenamefont {Resta}(1994)}]{RevModPhys.66.899}%
  \BibitemOpen
  \bibfield  {author} {\bibinfo {author} {\bibfnamefont {R.}~\bibnamefont
  {Resta}},\ }\href {http://link.aps.org/doi/10.1103/RevModPhys.66.899}
  {\bibfield  {journal} {\bibinfo  {journal} {Rev. Mod. Phys.}\ }\textbf
  {\bibinfo {volume} {66}},\ \bibinfo {pages} {899} (\bibinfo {year}
  {1994})}\BibitemShut {NoStop}%
\bibitem [{\citenamefont {Nakajima}\ \emph {et~al.}(2016)\citenamefont
  {Nakajima}, \citenamefont {Tomita}, \citenamefont {Taie}, \citenamefont
  {Ichinose}, \citenamefont {Ozawa}, \citenamefont {Wang}, \citenamefont
  {Troyer},\ and\ \citenamefont {Takahashi}}]{Nakajima}%
  \BibitemOpen
  \bibfield  {author} {\bibinfo {author} {\bibfnamefont {S.}~\bibnamefont
  {Nakajima}}, \bibinfo {author} {\bibfnamefont {T.}~\bibnamefont {Tomita}},
  \bibinfo {author} {\bibfnamefont {S.}~\bibnamefont {Taie}}, \bibinfo {author}
  {\bibfnamefont {T.}~\bibnamefont {Ichinose}}, \bibinfo {author}
  {\bibfnamefont {H.}~\bibnamefont {Ozawa}}, \bibinfo {author} {\bibfnamefont
  {L.}~\bibnamefont {Wang}}, \bibinfo {author} {\bibfnamefont {M.}~\bibnamefont
  {Troyer}}, \ and\ \bibinfo {author} {\bibfnamefont {Y.}~\bibnamefont
  {Takahashi}},\ }\href {http://dx.doi.org/10.1038/nphys3622} {\bibfield
  {journal} {\bibinfo  {journal} {Nat Phys}\ }\textbf {\bibinfo {volume}
  {12}},\ \bibinfo {pages} {296} (\bibinfo {year} {2016})}\BibitemShut
  {NoStop}%
\bibitem [{\citenamefont {Lohse}\ \emph {et~al.}(2016)\citenamefont {Lohse},
  \citenamefont {Schweizer}, \citenamefont {Zilberberg}, \citenamefont
  {Aidelsburger},\ and\ \citenamefont {Bloch}}]{Lohse}%
  \BibitemOpen
  \bibfield  {author} {\bibinfo {author} {\bibfnamefont {M.}~\bibnamefont
  {Lohse}}, \bibinfo {author} {\bibfnamefont {C.}~\bibnamefont {Schweizer}},
  \bibinfo {author} {\bibfnamefont {O.}~\bibnamefont {Zilberberg}}, \bibinfo
  {author} {\bibfnamefont {M.}~\bibnamefont {Aidelsburger}}, \ and\ \bibinfo
  {author} {\bibfnamefont {I.}~\bibnamefont {Bloch}},\ }\href
  {http://dx.doi.org/10.1038/nphys3584} {\bibfield  {journal} {\bibinfo
  {journal} {Nat Phys}\ }\textbf {\bibinfo {volume} {12}},\ \bibinfo {pages}
  {350} (\bibinfo {year} {2016})}\BibitemShut {NoStop}%
\bibitem [{\citenamefont {Hayata}\ and\ \citenamefont
  {Hidaka}(2017{\natexlab{a}})}]{PhysRevB.95.125137}%
  \BibitemOpen
  \bibfield  {author} {\bibinfo {author} {\bibfnamefont {T.}~\bibnamefont
  {Hayata}}\ and\ \bibinfo {author} {\bibfnamefont {Y.}~\bibnamefont
  {Hidaka}},\ }\href {https://link.aps.org/doi/10.1103/PhysRevB.95.125137}
  {\bibfield  {journal} {\bibinfo  {journal} {Phys. Rev. B}\ }\textbf {\bibinfo
  {volume} {95}},\ \bibinfo {pages} {125137} (\bibinfo {year}
  {2017}{\natexlab{a}})}\BibitemShut {NoStop}%
\bibitem [{\citenamefont {Bychkov}\ and\ \citenamefont
  {Rashba}(1984{\natexlab{a}})}]{Rashba}%
  \BibitemOpen
  \bibfield  {author} {\bibinfo {author} {\bibfnamefont {Y.~A.}\ \bibnamefont
  {Bychkov}}\ and\ \bibinfo {author} {\bibfnamefont {E.~I.}\ \bibnamefont
  {Rashba}},\ }\href {http://www.jetpletters.ac.ru/ps/1264/article_19121.shtml}
  {\bibfield  {journal} {\bibinfo  {journal} {JETP Lett.}\ }\textbf {\bibinfo
  {volume} {39}},\ \bibinfo {pages} {78} (\bibinfo {year}
  {1984}{\natexlab{a}})}\BibitemShut {NoStop}%
\bibitem [{\citenamefont {Bychkov}\ and\ \citenamefont
  {Rashba}(1984{\natexlab{b}})}]{0022-3719-17-33-015}%
  \BibitemOpen
  \bibfield  {author} {\bibinfo {author} {\bibfnamefont {Y.~A.}\ \bibnamefont
  {Bychkov}}\ and\ \bibinfo {author} {\bibfnamefont {E.~I.}\ \bibnamefont
  {Rashba}},\ }\href {http://stacks.iop.org/0022-3719/17/i=33/a=015} {\bibfield
   {journal} {\bibinfo  {journal} {Journal of Physics C: Solid State Physics}\
  }\textbf {\bibinfo {volume} {17}},\ \bibinfo {pages} {6039} (\bibinfo {year}
  {1984}{\natexlab{b}})}\BibitemShut {NoStop}%
\bibitem [{\citenamefont {Oguchi}\ and\ \citenamefont
  {Shishidou}(2009)}]{0953-8984-21-9-092001}%
  \BibitemOpen
  \bibfield  {author} {\bibinfo {author} {\bibfnamefont {T.}~\bibnamefont
  {Oguchi}}\ and\ \bibinfo {author} {\bibfnamefont {T.}~\bibnamefont
  {Shishidou}},\ }\href {http://stacks.iop.org/0953-8984/21/i=9/a=092001}
  {\bibfield  {journal} {\bibinfo  {journal} {Journal of Physics: Condensed
  Matter}\ }\textbf {\bibinfo {volume} {21}},\ \bibinfo {pages} {092001}
  (\bibinfo {year} {2009})}\BibitemShut {NoStop}%
\bibitem [{\citenamefont {Simon}\ \emph {et~al.}(2010)\citenamefont {Simon},
  \citenamefont {Szilva}, \citenamefont {Ujfalussy}, \citenamefont
  {Lazarovits}, \citenamefont {Zarand},\ and\ \citenamefont
  {Szunyogh}}]{PhysRevB.81.235438}%
  \BibitemOpen
  \bibfield  {author} {\bibinfo {author} {\bibfnamefont {E.}~\bibnamefont
  {Simon}}, \bibinfo {author} {\bibfnamefont {A.}~\bibnamefont {Szilva}},
  \bibinfo {author} {\bibfnamefont {B.}~\bibnamefont {Ujfalussy}}, \bibinfo
  {author} {\bibfnamefont {B.}~\bibnamefont {Lazarovits}}, \bibinfo {author}
  {\bibfnamefont {G.}~\bibnamefont {Zarand}}, \ and\ \bibinfo {author}
  {\bibfnamefont {L.}~\bibnamefont {Szunyogh}},\ }\href
  {https://link.aps.org/doi/10.1103/PhysRevB.81.235438} {\bibfield  {journal}
  {\bibinfo  {journal} {Phys. Rev. B}\ }\textbf {\bibinfo {volume} {81}},\
  \bibinfo {pages} {235438} (\bibinfo {year} {2010})}\BibitemShut {NoStop}%
\bibitem [{\citenamefont {Nitta}\ \emph {et~al.}(1997)\citenamefont {Nitta},
  \citenamefont {Akazaki}, \citenamefont {Takayanagi},\ and\ \citenamefont
  {Enoki}}]{PhysRevLett.78.1335}%
  \BibitemOpen
  \bibfield  {author} {\bibinfo {author} {\bibfnamefont {J.}~\bibnamefont
  {Nitta}}, \bibinfo {author} {\bibfnamefont {T.}~\bibnamefont {Akazaki}},
  \bibinfo {author} {\bibfnamefont {H.}~\bibnamefont {Takayanagi}}, \ and\
  \bibinfo {author} {\bibfnamefont {T.}~\bibnamefont {Enoki}},\ }\href
  {https://link.aps.org/doi/10.1103/PhysRevLett.78.1335} {\bibfield  {journal}
  {\bibinfo  {journal} {Phys. Rev. Lett.}\ }\textbf {\bibinfo {volume} {78}},\
  \bibinfo {pages} {1335} (\bibinfo {year} {1997})}\BibitemShut {NoStop}%
\bibitem [{\citenamefont {Schliemann}\ and\ \citenamefont
  {Loss}(2003)}]{PhysRevB.68.165311}%
  \BibitemOpen
  \bibfield  {author} {\bibinfo {author} {\bibfnamefont {J.}~\bibnamefont
  {Schliemann}}\ and\ \bibinfo {author} {\bibfnamefont {D.}~\bibnamefont
  {Loss}},\ }\href {https://link.aps.org/doi/10.1103/PhysRevB.68.165311}
  {\bibfield  {journal} {\bibinfo  {journal} {Phys. Rev. B}\ }\textbf {\bibinfo
  {volume} {68}},\ \bibinfo {pages} {165311} (\bibinfo {year}
  {2003})}\BibitemShut {NoStop}%
\bibitem [{\citenamefont {Sinitsyn}\ \emph {et~al.}(2007)\citenamefont
  {Sinitsyn}, \citenamefont {MacDonald}, \citenamefont {Jungwirth},
  \citenamefont {Dugaev},\ and\ \citenamefont {Sinova}}]{PhysRevB.75.045315}%
  \BibitemOpen
  \bibfield  {author} {\bibinfo {author} {\bibfnamefont {N.~A.}\ \bibnamefont
  {Sinitsyn}}, \bibinfo {author} {\bibfnamefont {A.~H.}\ \bibnamefont
  {MacDonald}}, \bibinfo {author} {\bibfnamefont {T.}~\bibnamefont
  {Jungwirth}}, \bibinfo {author} {\bibfnamefont {V.~K.}\ \bibnamefont
  {Dugaev}}, \ and\ \bibinfo {author} {\bibfnamefont {J.}~\bibnamefont
  {Sinova}},\ }\href {https://link.aps.org/doi/10.1103/PhysRevB.75.045315}
  {\bibfield  {journal} {\bibinfo  {journal} {Phys. Rev. B}\ }\textbf {\bibinfo
  {volume} {75}},\ \bibinfo {pages} {045315} (\bibinfo {year}
  {2007})}\BibitemShut {NoStop}%
\bibitem [{\citenamefont {Ishizuka}\ \emph {et~al.}(2016)\citenamefont
  {Ishizuka}, \citenamefont {Hayata}, \citenamefont {Ueda},\ and\ \citenamefont
  {Nagaosa}}]{PhysRevLett.117.216601}%
  \BibitemOpen
  \bibfield  {author} {\bibinfo {author} {\bibfnamefont {H.}~\bibnamefont
  {Ishizuka}}, \bibinfo {author} {\bibfnamefont {T.}~\bibnamefont {Hayata}},
  \bibinfo {author} {\bibfnamefont {M.}~\bibnamefont {Ueda}}, \ and\ \bibinfo
  {author} {\bibfnamefont {N.}~\bibnamefont {Nagaosa}},\ }\href
  {https://link.aps.org/doi/10.1103/PhysRevLett.117.216601} {\bibfield
  {journal} {\bibinfo  {journal} {Phys. Rev. Lett.}\ }\textbf {\bibinfo
  {volume} {117}},\ \bibinfo {pages} {216601} (\bibinfo {year}
  {2016})}\BibitemShut {NoStop}%
\bibitem [{\citenamefont {{Ishizuka}}\ \emph {et~al.}(2017)\citenamefont
  {{Ishizuka}}, \citenamefont {{Hayata}}, \citenamefont {{Ueda}},\ and\
  \citenamefont {{Nagaosa}}}]{2017arXiv170201450I}%
  \BibitemOpen
  \bibfield  {author} {\bibinfo {author} {\bibfnamefont {H.}~\bibnamefont
  {{Ishizuka}}}, \bibinfo {author} {\bibfnamefont {T.}~\bibnamefont
  {{Hayata}}}, \bibinfo {author} {\bibfnamefont {M.}~\bibnamefont {{Ueda}}}, \
  and\ \bibinfo {author} {\bibfnamefont {N.}~\bibnamefont {{Nagaosa}}},\
  }\href@noop {} {\bibfield  {journal} {\bibinfo  {journal} {ArXiv e-prints}\ }
  (\bibinfo {year} {2017})},\ \Eprint {http://arxiv.org/abs/1702.01450}
  {arXiv:1702.01450 [cond-mat.mes-hall]} \BibitemShut {NoStop}%
\bibitem [{\citenamefont {Stephanov}\ and\ \citenamefont
  {Yin}(2012)}]{PhysRevLett.109.162001}%
  \BibitemOpen
  \bibfield  {author} {\bibinfo {author} {\bibfnamefont {M.~A.}\ \bibnamefont
  {Stephanov}}\ and\ \bibinfo {author} {\bibfnamefont {Y.}~\bibnamefont
  {Yin}},\ }\href {http://link.aps.org/doi/10.1103/PhysRevLett.109.162001}
  {\bibfield  {journal} {\bibinfo  {journal} {Phys. Rev. Lett.}\ }\textbf
  {\bibinfo {volume} {109}},\ \bibinfo {pages} {162001} (\bibinfo {year}
  {2012})}\BibitemShut {NoStop}%
\bibitem [{\citenamefont {Son}\ and\ \citenamefont
  {Yamamoto}(2012)}]{PhysRevLett.109.181602}%
  \BibitemOpen
  \bibfield  {author} {\bibinfo {author} {\bibfnamefont {D.~T.}\ \bibnamefont
  {Son}}\ and\ \bibinfo {author} {\bibfnamefont {N.}~\bibnamefont {Yamamoto}},\
  }\href {http://link.aps.org/doi/10.1103/PhysRevLett.109.181602} {\bibfield
  {journal} {\bibinfo  {journal} {Phys. Rev. Lett.}\ }\textbf {\bibinfo
  {volume} {109}},\ \bibinfo {pages} {181602} (\bibinfo {year}
  {2012})}\BibitemShut {NoStop}%
\bibitem [{\citenamefont {Chen}\ \emph {et~al.}(2013)\citenamefont {Chen},
  \citenamefont {Pu}, \citenamefont {Wang},\ and\ \citenamefont
  {Wang}}]{PhysRevLett.110.262301}%
  \BibitemOpen
  \bibfield  {author} {\bibinfo {author} {\bibfnamefont {J.-W.}\ \bibnamefont
  {Chen}}, \bibinfo {author} {\bibfnamefont {S.}~\bibnamefont {Pu}}, \bibinfo
  {author} {\bibfnamefont {Q.}~\bibnamefont {Wang}}, \ and\ \bibinfo {author}
  {\bibfnamefont {X.-N.}\ \bibnamefont {Wang}},\ }\href
  {http://link.aps.org/doi/10.1103/PhysRevLett.110.262301} {\bibfield
  {journal} {\bibinfo  {journal} {Phys. Rev. Lett.}\ }\textbf {\bibinfo
  {volume} {110}},\ \bibinfo {pages} {262301} (\bibinfo {year}
  {2013})}\BibitemShut {NoStop}%
\bibitem [{\citenamefont {Huang}\ \emph
  {et~al.}(2015{\natexlab{b}})\citenamefont {Huang}, \citenamefont {Xu},
  \citenamefont {Belopolski}, \citenamefont {Lee}, \citenamefont {Chang},
  \citenamefont {Wang}, \citenamefont {Alidoust}, \citenamefont {Bian},
  \citenamefont {Neupane}, \citenamefont {Zhang}, \citenamefont {Jia},
  \citenamefont {Bansil}, \citenamefont {Lin},\ and\ \citenamefont
  {Hasan}}]{Huang}%
  \BibitemOpen
  \bibfield  {author} {\bibinfo {author} {\bibfnamefont {S.-M.}\ \bibnamefont
  {Huang}}, \bibinfo {author} {\bibfnamefont {S.-Y.}\ \bibnamefont {Xu}},
  \bibinfo {author} {\bibfnamefont {I.}~\bibnamefont {Belopolski}}, \bibinfo
  {author} {\bibfnamefont {C.-C.}\ \bibnamefont {Lee}}, \bibinfo {author}
  {\bibfnamefont {G.}~\bibnamefont {Chang}}, \bibinfo {author} {\bibfnamefont
  {B.}~\bibnamefont {Wang}}, \bibinfo {author} {\bibfnamefont {N.}~\bibnamefont
  {Alidoust}}, \bibinfo {author} {\bibfnamefont {G.}~\bibnamefont {Bian}},
  \bibinfo {author} {\bibfnamefont {M.}~\bibnamefont {Neupane}}, \bibinfo
  {author} {\bibfnamefont {C.}~\bibnamefont {Zhang}}, \bibinfo {author}
  {\bibfnamefont {S.}~\bibnamefont {Jia}}, \bibinfo {author} {\bibfnamefont
  {A.}~\bibnamefont {Bansil}}, \bibinfo {author} {\bibfnamefont
  {H.}~\bibnamefont {Lin}}, \ and\ \bibinfo {author} {\bibfnamefont {M.~Z.}\
  \bibnamefont {Hasan}},\ }\href {http://dx.doi.org/10.1038/ncomms8373}
  {\bibfield  {journal} {\bibinfo  {journal} {Nature Communications}\ }\textbf
  {\bibinfo {volume} {6}},\ \bibinfo {pages} {7373 EP } (\bibinfo {year}
  {2015}{\natexlab{b}})}\BibitemShut {NoStop}%
\bibitem [{\citenamefont {Weng}\ \emph {et~al.}(2015)\citenamefont {Weng},
  \citenamefont {Fang}, \citenamefont {Fang}, \citenamefont {Bernevig},\ and\
  \citenamefont {Dai}}]{PhysRevX.5.011029}%
  \BibitemOpen
  \bibfield  {author} {\bibinfo {author} {\bibfnamefont {H.}~\bibnamefont
  {Weng}}, \bibinfo {author} {\bibfnamefont {C.}~\bibnamefont {Fang}}, \bibinfo
  {author} {\bibfnamefont {Z.}~\bibnamefont {Fang}}, \bibinfo {author}
  {\bibfnamefont {B.~A.}\ \bibnamefont {Bernevig}}, \ and\ \bibinfo {author}
  {\bibfnamefont {X.}~\bibnamefont {Dai}},\ }\href
  {https://link.aps.org/doi/10.1103/PhysRevX.5.011029} {\bibfield  {journal}
  {\bibinfo  {journal} {Phys. Rev. X}\ }\textbf {\bibinfo {volume} {5}},\
  \bibinfo {pages} {011029} (\bibinfo {year} {2015})}\BibitemShut {NoStop}%
\bibitem [{\citenamefont {Lv}\ \emph {et~al.}(2015)\citenamefont {Lv},
  \citenamefont {Weng}, \citenamefont {Fu}, \citenamefont {Wang}, \citenamefont
  {Miao}, \citenamefont {Ma}, \citenamefont {Richard}, \citenamefont {Huang},
  \citenamefont {Zhao}, \citenamefont {Chen}, \citenamefont {Fang},
  \citenamefont {Dai}, \citenamefont {Qian},\ and\ \citenamefont
  {Ding}}]{PhysRevX.5.031013}%
  \BibitemOpen
  \bibfield  {author} {\bibinfo {author} {\bibfnamefont {B.~Q.}\ \bibnamefont
  {Lv}}, \bibinfo {author} {\bibfnamefont {H.~M.}\ \bibnamefont {Weng}},
  \bibinfo {author} {\bibfnamefont {B.~B.}\ \bibnamefont {Fu}}, \bibinfo
  {author} {\bibfnamefont {X.~P.}\ \bibnamefont {Wang}}, \bibinfo {author}
  {\bibfnamefont {H.}~\bibnamefont {Miao}}, \bibinfo {author} {\bibfnamefont
  {J.}~\bibnamefont {Ma}}, \bibinfo {author} {\bibfnamefont {P.}~\bibnamefont
  {Richard}}, \bibinfo {author} {\bibfnamefont {X.~C.}\ \bibnamefont {Huang}},
  \bibinfo {author} {\bibfnamefont {L.~X.}\ \bibnamefont {Zhao}}, \bibinfo
  {author} {\bibfnamefont {G.~F.}\ \bibnamefont {Chen}}, \bibinfo {author}
  {\bibfnamefont {Z.}~\bibnamefont {Fang}}, \bibinfo {author} {\bibfnamefont
  {X.}~\bibnamefont {Dai}}, \bibinfo {author} {\bibfnamefont {T.}~\bibnamefont
  {Qian}}, \ and\ \bibinfo {author} {\bibfnamefont {H.}~\bibnamefont {Ding}},\
  }\href {https://link.aps.org/doi/10.1103/PhysRevX.5.031013} {\bibfield
  {journal} {\bibinfo  {journal} {Phys. Rev. X}\ }\textbf {\bibinfo {volume}
  {5}},\ \bibinfo {pages} {031013} (\bibinfo {year} {2015})}\BibitemShut
  {NoStop}%
\bibitem [{\citenamefont {Xu}\ \emph {et~al.}(2015)\citenamefont {Xu},
  \citenamefont {Belopolski}, \citenamefont {Alidoust}, \citenamefont
  {Neupane}, \citenamefont {Bian}, \citenamefont {Zhang}, \citenamefont
  {Sankar}, \citenamefont {Chang}, \citenamefont {Yuan}, \citenamefont {Lee},
  \citenamefont {Huang}, \citenamefont {Zheng}, \citenamefont {Ma},
  \citenamefont {Sanchez}, \citenamefont {Wang}, \citenamefont {Bansil},
  \citenamefont {Chou}, \citenamefont {Shibayev}, \citenamefont {Lin},
  \citenamefont {Jia},\ and\ \citenamefont {Hasan}}]{Xu613}%
  \BibitemOpen
  \bibfield  {author} {\bibinfo {author} {\bibfnamefont {S.-Y.}\ \bibnamefont
  {Xu}}, \bibinfo {author} {\bibfnamefont {I.}~\bibnamefont {Belopolski}},
  \bibinfo {author} {\bibfnamefont {N.}~\bibnamefont {Alidoust}}, \bibinfo
  {author} {\bibfnamefont {M.}~\bibnamefont {Neupane}}, \bibinfo {author}
  {\bibfnamefont {G.}~\bibnamefont {Bian}}, \bibinfo {author} {\bibfnamefont
  {C.}~\bibnamefont {Zhang}}, \bibinfo {author} {\bibfnamefont
  {R.}~\bibnamefont {Sankar}}, \bibinfo {author} {\bibfnamefont
  {G.}~\bibnamefont {Chang}}, \bibinfo {author} {\bibfnamefont
  {Z.}~\bibnamefont {Yuan}}, \bibinfo {author} {\bibfnamefont {C.-C.}\
  \bibnamefont {Lee}}, \bibinfo {author} {\bibfnamefont {S.-M.}\ \bibnamefont
  {Huang}}, \bibinfo {author} {\bibfnamefont {H.}~\bibnamefont {Zheng}},
  \bibinfo {author} {\bibfnamefont {J.}~\bibnamefont {Ma}}, \bibinfo {author}
  {\bibfnamefont {D.~S.}\ \bibnamefont {Sanchez}}, \bibinfo {author}
  {\bibfnamefont {B.}~\bibnamefont {Wang}}, \bibinfo {author} {\bibfnamefont
  {A.}~\bibnamefont {Bansil}}, \bibinfo {author} {\bibfnamefont
  {F.}~\bibnamefont {Chou}}, \bibinfo {author} {\bibfnamefont {P.~P.}\
  \bibnamefont {Shibayev}}, \bibinfo {author} {\bibfnamefont {H.}~\bibnamefont
  {Lin}}, \bibinfo {author} {\bibfnamefont {S.}~\bibnamefont {Jia}}, \ and\
  \bibinfo {author} {\bibfnamefont {M.~Z.}\ \bibnamefont {Hasan}},\ }\href
  {http://science.sciencemag.org/content/349/6248/613} {\bibfield  {journal}
  {\bibinfo  {journal} {Science}\ }\textbf {\bibinfo {volume} {349}},\ \bibinfo
  {pages} {613} (\bibinfo {year} {2015})}\BibitemShut {NoStop}%
\bibitem [{\citenamefont {Huang}\ \emph {et~al.}(2016)\citenamefont {Huang},
  \citenamefont {Xu}, \citenamefont {Belopolski}, \citenamefont {Lee},
  \citenamefont {Chang}, \citenamefont {Chang}, \citenamefont {Wang},
  \citenamefont {Alidoust}, \citenamefont {Bian}, \citenamefont {Neupane},
  \citenamefont {Sanchez}, \citenamefont {Zheng}, \citenamefont {Jeng},
  \citenamefont {Bansil}, \citenamefont {Neupert}, \citenamefont {Lin},\ and\
  \citenamefont {Hasan}}]{Huang02022016}%
  \BibitemOpen
  \bibfield  {author} {\bibinfo {author} {\bibfnamefont {S.-M.}\ \bibnamefont
  {Huang}}, \bibinfo {author} {\bibfnamefont {S.-Y.}\ \bibnamefont {Xu}},
  \bibinfo {author} {\bibfnamefont {I.}~\bibnamefont {Belopolski}}, \bibinfo
  {author} {\bibfnamefont {C.-C.}\ \bibnamefont {Lee}}, \bibinfo {author}
  {\bibfnamefont {G.}~\bibnamefont {Chang}}, \bibinfo {author} {\bibfnamefont
  {T.-R.}\ \bibnamefont {Chang}}, \bibinfo {author} {\bibfnamefont
  {B.}~\bibnamefont {Wang}}, \bibinfo {author} {\bibfnamefont {N.}~\bibnamefont
  {Alidoust}}, \bibinfo {author} {\bibfnamefont {G.}~\bibnamefont {Bian}},
  \bibinfo {author} {\bibfnamefont {M.}~\bibnamefont {Neupane}}, \bibinfo
  {author} {\bibfnamefont {D.}~\bibnamefont {Sanchez}}, \bibinfo {author}
  {\bibfnamefont {H.}~\bibnamefont {Zheng}}, \bibinfo {author} {\bibfnamefont
  {H.-T.}\ \bibnamefont {Jeng}}, \bibinfo {author} {\bibfnamefont
  {A.}~\bibnamefont {Bansil}}, \bibinfo {author} {\bibfnamefont
  {T.}~\bibnamefont {Neupert}}, \bibinfo {author} {\bibfnamefont
  {H.}~\bibnamefont {Lin}}, \ and\ \bibinfo {author} {\bibfnamefont {M.~Z.}\
  \bibnamefont {Hasan}},\ }\href
  {http://www.pnas.org/content/113/5/1180.abstract} {\bibfield  {journal}
  {\bibinfo  {journal} {Proceedings of the National Academy of Sciences}\
  }\textbf {\bibinfo {volume} {113}},\ \bibinfo {pages} {1180} (\bibinfo {year}
  {2016})}\BibitemShut {NoStop}%
\bibitem [{\citenamefont {Hoyos}\ and\ \citenamefont
  {Son}(2012)}]{PhysRevLett.108.066805}%
  \BibitemOpen
  \bibfield  {author} {\bibinfo {author} {\bibfnamefont {C.}~\bibnamefont
  {Hoyos}}\ and\ \bibinfo {author} {\bibfnamefont {D.~T.}\ \bibnamefont
  {Son}},\ }\href {https://link.aps.org/doi/10.1103/PhysRevLett.108.066805}
  {\bibfield  {journal} {\bibinfo  {journal} {Phys. Rev. Lett.}\ }\textbf
  {\bibinfo {volume} {108}},\ \bibinfo {pages} {066805} (\bibinfo {year}
  {2012})}\BibitemShut {NoStop}%
\bibitem [{\citenamefont {Scaffidi}\ \emph {et~al.}(2017)\citenamefont
  {Scaffidi}, \citenamefont {Nandi}, \citenamefont {Schmidt}, \citenamefont
  {Mackenzie},\ and\ \citenamefont {Moore}}]{PhysRevLett.118.226601}%
  \BibitemOpen
  \bibfield  {author} {\bibinfo {author} {\bibfnamefont {T.}~\bibnamefont
  {Scaffidi}}, \bibinfo {author} {\bibfnamefont {N.}~\bibnamefont {Nandi}},
  \bibinfo {author} {\bibfnamefont {B.}~\bibnamefont {Schmidt}}, \bibinfo
  {author} {\bibfnamefont {A.~P.}\ \bibnamefont {Mackenzie}}, \ and\ \bibinfo
  {author} {\bibfnamefont {J.~E.}\ \bibnamefont {Moore}},\ }\href
  {https://link.aps.org/doi/10.1103/PhysRevLett.118.226601} {\bibfield
  {journal} {\bibinfo  {journal} {Phys. Rev. Lett.}\ }\textbf {\bibinfo
  {volume} {118}},\ \bibinfo {pages} {226601} (\bibinfo {year}
  {2017})}\BibitemShut {NoStop}%
\bibitem [{\citenamefont {Morimoto}\ and\ \citenamefont
  {Nagaosa}(2016)}]{Morimotoe1501524}%
  \BibitemOpen
  \bibfield  {author} {\bibinfo {author} {\bibfnamefont {T.}~\bibnamefont
  {Morimoto}}\ and\ \bibinfo {author} {\bibfnamefont {N.}~\bibnamefont
  {Nagaosa}},\ }\href {http://advances.sciencemag.org/content/2/5/e1501524}
  {\bibfield  {journal} {\bibinfo  {journal} {Science Advances}\ }\textbf
  {\bibinfo {volume} {2}} (\bibinfo {year} {2016})}\BibitemShut {NoStop}%
\bibitem [{\citenamefont {Cortijo}(2016)}]{PhysRevB.94.235123}%
  \BibitemOpen
  \bibfield  {author} {\bibinfo {author} {\bibfnamefont {A.}~\bibnamefont
  {Cortijo}},\ }\href {https://link.aps.org/doi/10.1103/PhysRevB.94.235123}
  {\bibfield  {journal} {\bibinfo  {journal} {Phys. Rev. B}\ }\textbf {\bibinfo
  {volume} {94}},\ \bibinfo {pages} {235123} (\bibinfo {year}
  {2016})}\BibitemShut {NoStop}%
\bibitem [{\citenamefont {Morimoto}\ \emph {et~al.}(2016)\citenamefont
  {Morimoto}, \citenamefont {Zhong}, \citenamefont {Orenstein},\ and\
  \citenamefont {Moore}}]{PhysRevB.94.245121}%
  \BibitemOpen
  \bibfield  {author} {\bibinfo {author} {\bibfnamefont {T.}~\bibnamefont
  {Morimoto}}, \bibinfo {author} {\bibfnamefont {S.}~\bibnamefont {Zhong}},
  \bibinfo {author} {\bibfnamefont {J.}~\bibnamefont {Orenstein}}, \ and\
  \bibinfo {author} {\bibfnamefont {J.~E.}\ \bibnamefont {Moore}},\ }\href
  {https://link.aps.org/doi/10.1103/PhysRevB.94.245121} {\bibfield  {journal}
  {\bibinfo  {journal} {Phys. Rev. B}\ }\textbf {\bibinfo {volume} {94}},\
  \bibinfo {pages} {245121} (\bibinfo {year} {2016})}\BibitemShut {NoStop}%
\bibitem [{\citenamefont {Culcer}\ \emph {et~al.}(2005)\citenamefont {Culcer},
  \citenamefont {Yao},\ and\ \citenamefont {Niu}}]{PhysRevB.72.085110}%
  \BibitemOpen
  \bibfield  {author} {\bibinfo {author} {\bibfnamefont {D.}~\bibnamefont
  {Culcer}}, \bibinfo {author} {\bibfnamefont {Y.}~\bibnamefont {Yao}}, \ and\
  \bibinfo {author} {\bibfnamefont {Q.}~\bibnamefont {Niu}},\ }\href
  {https://link.aps.org/doi/10.1103/PhysRevB.72.085110} {\bibfield  {journal}
  {\bibinfo  {journal} {Phys. Rev. B}\ }\textbf {\bibinfo {volume} {72}},\
  \bibinfo {pages} {085110} (\bibinfo {year} {2005})}\BibitemShut {NoStop}%
\bibitem [{\citenamefont {Shindou}\ and\ \citenamefont
  {Imura}(2005)}]{Shindou2005399}%
  \BibitemOpen
  \bibfield  {author} {\bibinfo {author} {\bibfnamefont {R.}~\bibnamefont
  {Shindou}}\ and\ \bibinfo {author} {\bibfnamefont {K.-I.}\ \bibnamefont
  {Imura}},\ }\href
  {http://www.sciencedirect.com/science/article/pii/S0550321305004360}
  {\bibfield  {journal} {\bibinfo  {journal} {Nuclear Physics B}\ }\textbf
  {\bibinfo {volume} {720}},\ \bibinfo {pages} {399 } (\bibinfo {year}
  {2005})}\BibitemShut {NoStop}%
\bibitem [{\citenamefont {Chang}\ and\ \citenamefont
  {Niu}(2008)}]{0953-8984-20-19-193202}%
  \BibitemOpen
  \bibfield  {author} {\bibinfo {author} {\bibfnamefont {M.-C.}\ \bibnamefont
  {Chang}}\ and\ \bibinfo {author} {\bibfnamefont {Q.}~\bibnamefont {Niu}},\
  }\href {http://stacks.iop.org/0953-8984/20/i=19/a=193202} {\bibfield
  {journal} {\bibinfo  {journal} {Journal of Physics: Condensed Matter}\
  }\textbf {\bibinfo {volume} {20}},\ \bibinfo {pages} {193202} (\bibinfo
  {year} {2008})}\BibitemShut {NoStop}%
\bibitem [{\citenamefont {Hayata}\ and\ \citenamefont
  {Hidaka}(2017{\natexlab{b}})}]{Hayata:2017ihy}%
  \BibitemOpen
  \bibfield  {author} {\bibinfo {author} {\bibfnamefont {T.}~\bibnamefont
  {Hayata}}\ and\ \bibinfo {author} {\bibfnamefont {Y.}~\bibnamefont
  {Hidaka}},\ }\href@noop {} {\bibfield  {journal} {\bibinfo  {journal} {ArXiv
  e-prints}\ } (\bibinfo {year} {2017}{\natexlab{b}})},\ \Eprint
  {http://arxiv.org/abs/1701.04012} {arXiv:1701.04012 [cond-mat.mes-hall]}
  \BibitemShut {NoStop}%
%%CITATION = ARXIV:1701.04012;%%
\end{thebibliography}%

\end{document}